\documentclass[10pt,twocolumn,letterpaper]{article}

\usepackage{cvpr}
\usepackage{times}
\usepackage{epsfig}
\usepackage{graphicx}
\usepackage{amsmath}
\usepackage{amssymb}

\usepackage[breaklinks=true,bookmarks=false]{hyperref}

\cvprfinalcopy 

\def\cvprPaperID{100} 

\begin{document}

\title{MXR-U-Nets for Real Time Hyperspectral Reconstruction}

\author{Akash Palrecha\\
Pixxel (Numancia Aerospace Limited)\\
Workbench Projects, Bangalore\\
{\tt\small akash@pixxel.co.in}
\and
Atmadeep Banerjee\\
Pixxel (Numancia Aerospace Limited)\\
Workbench Projects, Bangalore\\
{\tt\small atmadeep@pixxel.co.in}
}

\maketitle

\begin{abstract}
   In recent times, CNNs have made significant contributions to applications in image generation, super-resolution and style transfer. In this paper, we build upon the work of Howard and Gugger \cite{FastAI}, He et al. \cite{BagOfTricks} and Misra, D. \cite{Mish} and propose a CNN architecture that accurately reconstructs hyperspectral images from their RGB counterparts. We also propose a much shallower version of our best model with a $10\%$ relative memory footprint and $3x$ faster inference thus enabling real time video applications while still experiencing only about a $0.5\%$ decrease in performance. The implementation is available \href{https://github.com/akashpalrecha/hyperspectral-reconstruction}{\textbf{here}}.
\end{abstract}

\section{Introduction}

Hyperspectral imagery captures information about objects across a wide range of the electromagnetic spectrum. These images possess much more amount of useful information compared to standard RGB images and are especially useful in fields like remote sensing and medical diagnosis. The main problem with hyperspectral imagery is the expensive hardware required to capture these images, leading to a lack of availability of datasets in the public domain. RGB images, on the other hand, are easy to obtain. A system for accurate reconstruction of spectral bands of an RGB image would be beneficial for furthering research into the applications of hyperspectral images.

It may seem problematic to try to convert RGB images to hyperspectral images since the task essentially requires the generation of information that was never captured by an RGB camera, but, hyperspectral image pixel values have a strong correlation \cite{correlation} to their RGB counterparts. It is, therefore, possible, to learn a mapping from RGB to hyperspectral images, given enough data. In fact, from visual inspection of the quality of results obtained even with simple CNN based approaches, we believe this is an easier task than an analogous task of converting gray scale images to RGB.

Our approach consists of using a CNN to convert RGB to hyperspectral images. We use a U-Net \cite{Unet} based model with several key improvements taken from recent advancements in the fields of image generation, super-resolution and style transfer. We use an XResnet model, as proposed by He et al. \cite{BagOfTricks} (referred to as \textit{Resnet-D} in \cite{BagOfTricks}) with Mish \cite{Mish} activation function (replacing ReLU \cite{relu}) as the encoder. In the decoder, we use sub-pixel convolutions \cite{PixelShuffle} for upsampling. Finally, we incorporate blur layers(approach B in \cite{Blur}) and a self-attention layer from \cite{SAGAN} in our decoder. We adapted the general decoder architecture from Howard and Gugger's work in the Fastai \cite{FastAI} library and Antic's work in DeOldify \cite{DeOldify}. We study the effect of adding each component on the accuracy and running time. We also study the effect of changing the depth of the encoder. Our proposed family of architectures are capable of accurately reconstructing spectral bands from RGB images with very low inference times, on standard single GPU systems. This paper is describes a soltuion to the challenge posed in \cite{Arad_2020_CVPR_Workshops}

\section{Related Work}
The problem of spectral reconstruction of RGB images is an area of computer vision that has not been studied too extensively. Current state-of-the-art approaches are mostly CNN based. Older methods made use of sparse coding \cite{sparse}, but recent advances in CNNs and availability of relatively larger datasets for hyperspectral reconstruction have led to increasing research into neural network based approaches. Earlier CNN based approaches used relatively shallow networks \cite{shallow} or even hybrid approaches combining sparse-coding and neural networks \cite{hybridSparse}. The NTIRE 2018 spectral reconstruction challenge introduced the BGU hyperspectral dataset, which was much larger than existing datasets. The challenge saw various deep CNN based \cite{ntire2018} and a few GAN based approaches. The new state of the art was achieved by Shi et al. \cite{hscnn}. Baran and Timofte also did important work for lightweight real time spectral reconstruction in \cite{RaduEfficientReconstruct}

Our model is a modified version of the U-Net \cite{Unet} architecture. The U-Net is a popular deep learning architecture originally introduced to perform image segmentation. It has since then been used for a wide range of image-to-image tasks. It improves over standard encoder-decoder architectures by incorporating skip connections between the encoder and decoder, allowing much deeper models to be trained. Over time the original U-Net architecture has seen several key improvements. These include incorporating skip connections in the encoder \cite{linknet}, using sub-pixel convolutions in the decoder \cite{PixelShuffle} and using dilated convolution \cite{dlinknet}.

Our work is significantly inspired by Antic, J.'s work\cite{DeOldify} in reconstructing RGB bands from grayscale images. We use a modified version of perceptual loss \cite{Perceptual} in our network. This kind of loss function has proved useful in style-transfer \cite{Perceptual} and super-resolution \cite{SRGAN} applications. It makes networks focus on \textit{perceptual details} in an image. These details are not easily captured by standard evaluation metrics like RMSE, PSNR or MRAE but are readily visible to humans. We make use of sub-pixel convolutions \cite{PixelShuffle} for upsampling, in our decoder. It is an alternative to deconvolution operation for learned upsampling and is extensively used in super-resolution applications. It performs the convolution in a low resolution space and upsamples the result, instead of upsampling first. This approach is much more efficient while being mathematically equivalent to deconvolution.


\section{Method}
\subsection{Architecture}
Our proposed architecture uses models from the XResnet family with Mish activation function as the encoder(this architecture will be referred to as \textit{mxresnet} \cite{mxresnet}) in the U-Net. Skip connections between the encoder and decoder are made at four positions where the encoder subsamples the image. The final encoder output is passed through two successive convolutions and fed into the decoder.

\begin{figure}[t]
\begin{center}
   \includegraphics[width=0.5\linewidth]{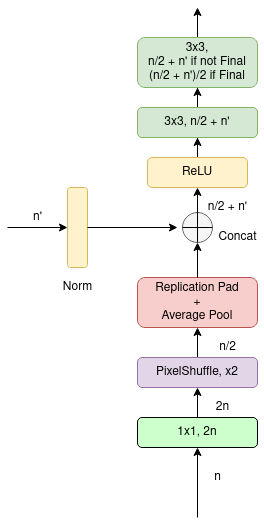}
\end{center}
   \caption{A U-Net Block}
\label{fig:unetBlock}
\label{fig:unetBlockOuter}
\end{figure}

The decoder consists of 4 upsampling blocks, each of which receives two input tensors and produces one output. The input from the previous decoder block is $2$x upsampled with a sub-pixel convolution with an ICNR\cite{ICNR} initialization scheme. A sub-pixel convolution operation combined with ICNR initialization has been attributed to performing high quality, checkerboard artifact free super-resolution. The upsampling is followed by a blur \cite{Blur} layer which consists of average pooling with a $2$x$2$ filter and stride of $1$. This operation also aims to reduce artifacts in generated images. The upsampled feature map is concatenated with the second input, which comes from an encoder skip connection. The final output is formed by passing the concatenated feature map through two successive convolutions. The second decoder block is followed by a self-attention layer as proposed by Zhang et al. in \cite{SAGAN}. This layer helps the network to focus on the relevant parts of the image. 

The decoder output is $2$x upsampled to make the resolution the same as the input image. This feature map is concatenated with the original RGB image and passed through a standard \textit{Resnet} block. We find that this concatenation operation provides significant improvements to our results. Finally, a $1$x$1$ convolution is used to bring down the channels to the desired number.
 

\subsection{Loss}
We use a slightly modified version of the loss function described by Johnson et al. in \cite{Perceptual}. Perceptual loss refer to a loss function that calculates the amount of dissimilarity between a generated image and the ground truth, based on perceptually relevant characteristics. Our loss function is the weighted sum of feature reconstruction loss, style reconstruction loss and pixel loss.
\begin{equation}
\begin{split}
  \ell = \sum\limits_{j}\alpha_{j}. \ell_{feat}^{\phi,j}(\hat y, y) + \\
         \sum\limits_{j}\beta_{j}. \ell_{style}^{\phi,j}(\hat y, y) + \\
         \gamma . \ell_{pixel}(\hat y, y)
\end{split}
\end{equation}

\noindent\textbf{Feature Reconstruction Loss}. We use a VGG16 \cite{vgg} network pretrained on Imagenet \cite{Imagenet} (also called the loss network) to compute the features of the model outputs. We modify the first layer of the network to contain filters with $31$ channels by copying over weights of the existing $3$ channels. Let $\phi_j(x)$ be the activations of the $j$th layer of the network $\phi$ when processing the image $x$; if $j$ is a convolutional layer then $\phi_j(x)$ will be a feature map of shape $C_j \times H_j \times W_j$. The \emph{feature reconstruction loss} is the mean L1 distance between feature representations:
\begin{equation}
  \ell_{feat}^{\phi,j}(\hat y, y) = 
  \frac1{C_jH_jW_j}\|\phi_j(\hat y) - \phi_j(y)\|
\end{equation}

We use the activations before the second, third and fourth max pool layers in the loss network to calculate our feature reconstruction loss.

\begin{figure}[t]
\begin{center}
   \includegraphics[width=0.6\linewidth]{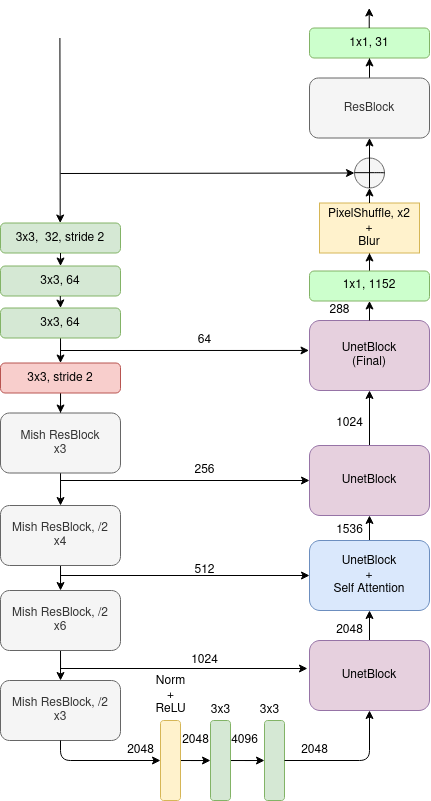}
\end{center}
   \caption{The mxresnet50 model}
\label{fig:arch}
\label{fig:archOuter}
\end{figure}

\noindent\textbf{Style Reconstruction Loss}
This loss was proposed by Gatys et al. in \cite{StyleTransfer} and adapted in \cite{Perceptual}. It constitutes calculating the Gram matrices of the loss network activations for the output and target images. The \emph{modified style reconstruction loss} is then the mean absolute difference between the Gram matrices of the output and target images. The Gram matrix can be computed efficiently by reshaping $\phi_j(x)$ into a matrix $\psi$ of shape $C_j\times H_jW_j$; then $G^\phi_j(x) = \psi\psi^T/C_jH_jW_j$.
\begin{equation}
  \ell_{style}^{\phi, j}(\hat y, y) = \|G^\phi_j(\hat y) - G^\phi_j(y)\|
\end{equation}

\begin{figure*}
  \centering \includegraphics[width=1.0\textwidth]{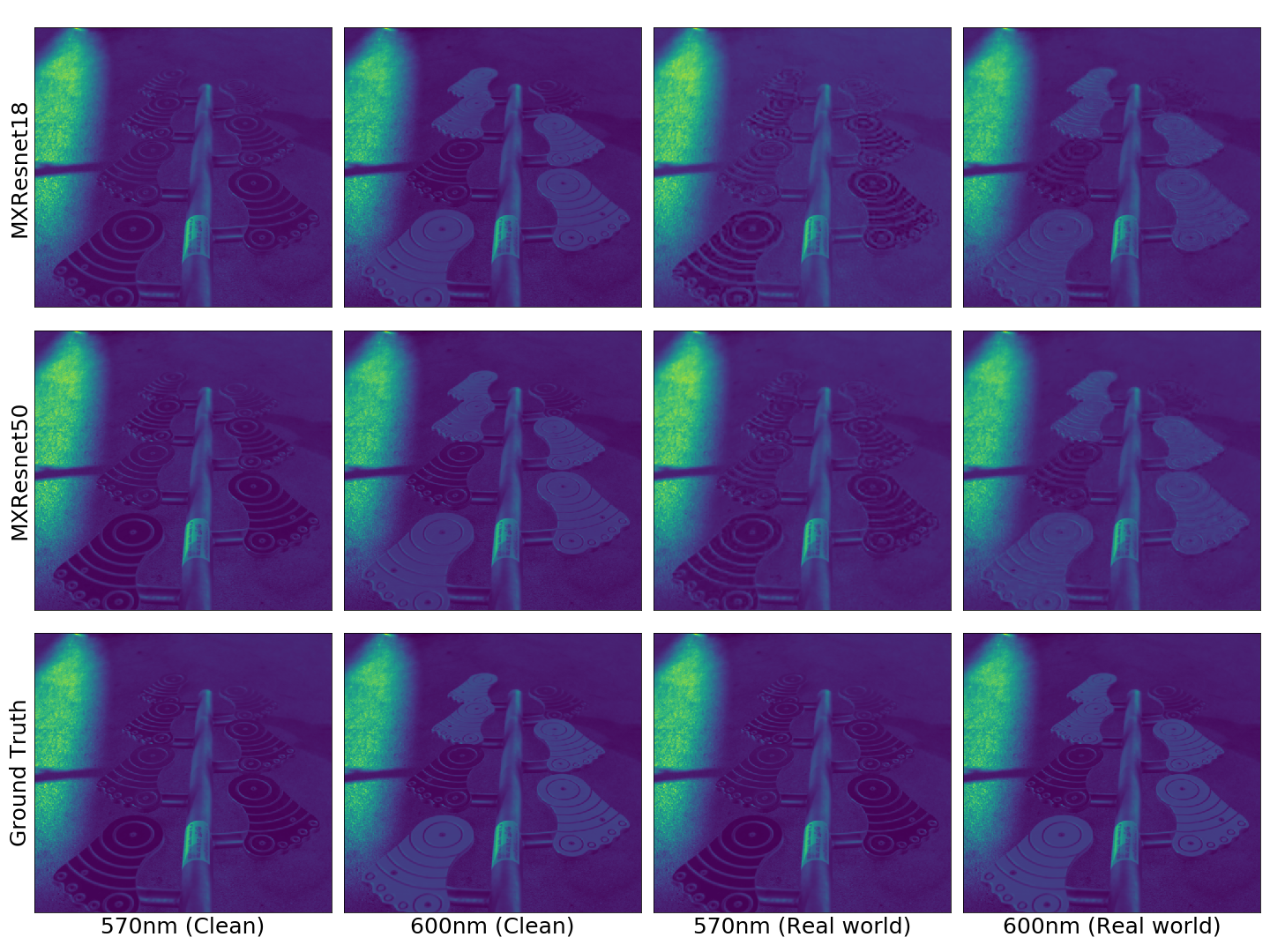}
  \caption{This figure visualizes the model outputs of \textit{mxresnet18} U-Net(our smallest model) and \textit{mxresnet50} U-Net (our largest model) on a Validation Image. On the Clean track, model outputs are visually indistinguishable from each other and the ground truth. On the Real World images, however, the differences are more clearly visible. The larger model produces visibly cleaner outputs.}
\end{figure*}

\noindent\textbf{Pixel Loss}
The \emph{pixel loss} is the mean Euclidean distance between the output image $\hat y$ and the target $y$. If both have shape $C\times H\times W$, then the pixel loss is defined as 
\begin{equation}
    \ell_{pixel}(\hat y, y) = \|\hat y - y\|^2_2 / CHW
\end{equation}

\section{Training}
We normalize the images and use the following data augmentation techniques: random flipping, random rotations, brightness and contrast jitter. All networks are trained for 200 epochs using the AdamW\cite{adamw} optimizer(Adam with weight decay) with a weight decay of 1e-3. The training follows the One Cycle schedule\cite{OneCycle}. Under this schedule, the learning rate is started at 1e-5 and increased to 1e-3 over 60 epochs following a half cosine curve. After the learning rate peaks, it is reduced to 1e-9 over another 140 epochs following a similar half cosine curve. The momentum of the optimizer goes through a similar but opposite cycle. It starts at 0.95 and is reduced to 0.85 over 60 epochs and again increased to 0.95 over 140 epochs. We also use mixed-precision training\cite{MixedPrecision} to lower training time and memory requirements. A single V100 GPU was used for all training runs. The entire training schedule takes $1.79$ hours ($43$s per epoch) for an \textit{mxresnet34} encoder based U-Net.

\begin{figure*}
  \centering \includegraphics[width=1.0\textwidth]{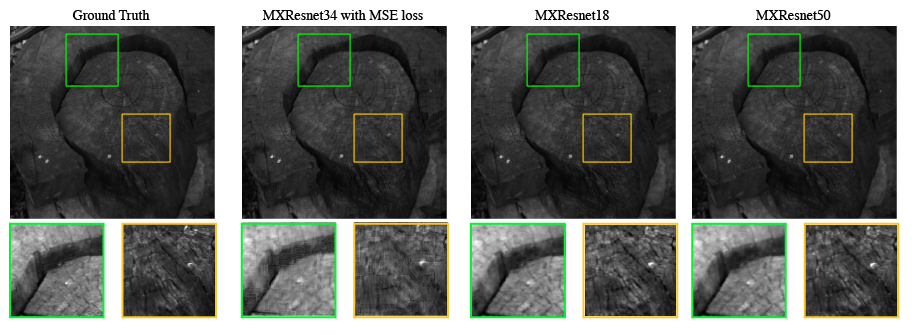}
  \caption{To demonstrate the effect of perceptual loss, we visualize the outputs of a model with an \textit{mxresnet34} encoder trained with \textit{MSE} loss, on a \textit{Real World} validation image, and compare it with the outputs of our models trained with perceptual loss. On zooming in, it can be seen that the \textit{MSE} model produces a considerably higher amount of artifacts.}
\end{figure*}

\begin{table}
\begin{center}
\begin{tabular}{l|l|r|}
\cline{2-3}
                                                                                                                       & \multicolumn{2}{c|}{\textbf{MRAE}}                            \\ \hline
\multicolumn{1}{|l|}{\textbf{Approach}}                                                                                & \textbf{Clean} & \textbf{Real World}              \\ \hline
\multicolumn{1}{|l|}{\textit{Resnet34 (pretrained)}}                                                                   & 0.055625             & 0.092532                               \\ \hline
\multicolumn{1}{|l|}{\textit{Resnet34 (no pretraining)}}                                                               & 0.052844             & 0.090132                               \\ \hline
\multicolumn{1}{|l|}{\textit{\begin{tabular}[c]{@{}l@{}}MXResnet34 \\ + Self Attention\\  + Blur\end{tabular}}}        & 0.052818             & 0.089132                               \\ \hline
\multicolumn{1}{|l|}{\textit{\begin{tabular}[c]{@{}l@{}}MXResnet34 \\ + Self Attention \\ with MSE Loss\end{tabular}}} & 0.169942             & 0.162509                               \\ \hline
\multicolumn{1}{|l|}{\textit{\begin{tabular}[c]{@{}l@{}}MXResnet18 \\ + Self Attention \\ + Blur\end{tabular}}}        & \textit{0.052089}    & \textit{0.088589}                      \\ \hline
\multicolumn{1}{|l|}{\textit{\begin{tabular}[c]{@{}l@{}}MXResnet50 \\ + Self Attention\\ + Blur\end{tabular}}}         & \textbf{0.045434}    & \textbf{0.083993} \\ \hline
\end{tabular}
\end{center}
\caption{Quantitative comparisons on both datasets for different approaches. Notable results are in bold or italics.}
\label{tab:1}
\end{table}

\section{Experiments}
Here we present some ablation studies comparing different variations of our proposed method. More specifically, we compare results for these encoder backbones: resnet34 \cite{Resnets}, mxresnet34 \cite{mxresnet}, mxresnet18 and mxresnet50. We also vary the presence of the self-attention layer along with the blur layer in the decoder networks. All networks have sub-pixel convolution layers in their decoders. The loss function for every experiment is the aforementioned perceptual loss combination unless otherwise specified.

\subsection{Dataset}
The dataset was provided in the \textit{New Trends in Image Restoration and Enhancement (NTIRE) Challenge on Spectral Reconstruction from RGB Images} at CVPR 2020 \cite{Arad_2020_CVPR_Workshops}. The datasets for both the competition tracks (\textit{Clean} and \textit{Real World}) consist of $450$ training images and $10$ validation images. The dataset for the \textit{clean} track of the competition consists of $8$-bit uncompressed RGB images and their $31$ channel hyperspectral counterparts as ground truth. For the \textit{real world} track, we have the JPEG compressed $8$-bit RGB images as the model input. In our experiments, the training and validation data for the models were as provided in the original datasets.

\subsection{Results}
As table \ref{tab:1} clearly indicates, a U-Net with an \textit{mxresnet50} encoder along with self-attention and blur in the decoder with perceptual losses produces the best results. We note that using a pre-trained model causes a slight reduction in performance as compared to other approaches that did not use any Imagenet \cite{Imagenet} pretraining. Further, adding self attention and blur to the decoder along with modifications to the original \textit{resnet} \cite{Resnets} architecture (\textit{xresnet} \cite{BagOfTricks} with a \textit{mish} \cite{Mish} activation function) gave a small performance boost.\\
Most surprisingly, though, an \textit{mxresnet18} based encoder was able to outperform all our other approaches except the \textit{mxresnet50} encoder while still being significantly shallow and computationally efficient as opposed to other approaches. 

\subsection{Inference Time}
Our model with the \textit{mxresnet50} encoder takes $0.159$ seconds per image during inference time. While all other approaches take $0.037$ to $0.042$ seconds per image during inference. Notably, the network with an \textit{mxresnet18} backbone takes $0.037$ seconds during inference making it suitable for real time video. The \textit{mxresnet18} encoder model has about $10$ times fewer parameters as compared to the \textit{mxresnet50} encoder model ($31.35$M vs $342.07$M) while still reducing the performance by only $0.006655$ and $0.004596$ on the \textit{clean} and \textit{real world} tracks with respect to the \textit{MRAE} metric.

\section{Conclusion}
In this paper, we use an encoder-decoder network based on the U-Net architecture with some of the recent advances/improvements in deep learning. We use the \textit{resnet-d}\cite{BagOfTricks} architecture with the \textit{mish} \cite{Mish} activation function as the encoder. In the decoder, we use sub-pixel convolution \cite{PixelShuffle} layers for upsampling to help increase efficiency, blur \cite{Blur} layers to reduce checkerboard artifacts and a self-attention layer \cite{SAGAN} to focus the network on finer details. All these improvements allow our approach to produce good results even with a relatively shallow encoder network such as the \textit{mxresnet18}. We note that much of these improvements were originally conceived and implemented in the FastAI \cite{FastAI} library by Howard, Gugger and Antic's contributions. Our contribution has been to combine the \textit{mxresnet} base architecture with these improvements that were made by the people mentioned above.
We also introduce a model based on the \textit{mxresnet18} encoder that is suitable for real-time video applications with an inference time of $0.037$ seconds without any significant drop in performance.

{\small
\bibliographystyle{ieee_fullname}
\bibliography{egbib}
}

\end{document}